\documentclass[pre,preprint, superscriptaddress, showpacs,amsmath,amssymb]{revtex4}
\usepackage{graphicx}
\usepackage{amsmath}
\usepackage{mathtools}
\usepackage[pdf]{pstricks}
\usepackage{epstopdf}

\begin{document}

\title{Analytical results regarding electrostatic resonances of surface 
phonon/plasmon polaritons: separation of variables with a twist }

\author{R. C. Voicu}
\affiliation{Research Centre for Integrated Systems, Nanotechnologies, and Carbon Based Materials,\\ 
National Institute for Research and Development in Microtechnologies-IMT, 126A, Erou Iancu Nicolae Street, 
Bucharest, ROMANIA}
\email{titus.sandu@imt.ro}
\author{T. Sandu}
\affiliation{Research Centre for Integrated Systems, Nanotechnologies, and Carbon Based Materials,\\ 
National Institute for Research and Development in Microtechnologies-IMT, 126A, Erou Iancu Nicolae Street, 
Bucharest, ROMANIA}

\date{\today}

\begin{abstract}
The boundary integral equation method ascertains explicit relations between localized surface phonon and plasmon polariton resonances and the eigenvalues of its associated electrostatic operator. We show that group-theoretical analysis of Laplace equation can be used to 
calculate the full set of eigenvalues and eigenfunctions of the electrostatic operator for shapes and shells 
described by separable coordinate systems. These results not only unify and generalize many existing studies but also offer the opportunity to expand the study of phenomena like cloaking by anomalous localized resonance. For that reason we calculate the eigenvalues and eigenfunctions of elliptic and circular cylinders. We illustrate the benefits of using the boundary integral equation method to interpret recent experiments involving localized surface phonon polariton resonances and the size scaling of plasmon resonances in graphene nano-disks. Finally, symmetry-based operator analysis can be extended from electrostatic to full-wave regime. Thus, bound states of light in the continuum can be studied for shapes beyond spherical configurations. 
\end{abstract}

\maketitle
\section{Introduction}
Materials with negative permittivity allow light confinement to sub-diffraction limit and field enhancement at the interface with 
ordinary dielectrics \cite{Barnes2003}. 
At optical frequencies noble metals 
exhibit such behaviour leading to numerous applications of the field 
called plasmonics \cite{Ozbay2006}. 
These applications include photonic circuits \cite{Ozbay2006} and 
biosensing \cite{Brolo2012}.
Metals have been promising since, with the advent of 
nanotechnology, a large variety of structures, shapes, and sizes can be 
tailored to obtain large tunability of plasmon resonances from ultraviolet 
(UV) to visible and near infrared (IR)  \cite{Noguez2007}.

Nowadays, plasmonics has reached the level of maturity such that new plasmonic materials 
are required for real applications \cite{Guler2015}. Plasmon excitations belong to 
the class of polaritons which are mixed states made from an elementary 
excitation (i.e., dipole-active such as a phonon, plasmon, magnon, exciton) 
coupled to a photon  \cite{Agranovich1982}. Hence, plasmon polaritons are mixed states made from 
collective excitations of free electrons in metals and photons, while phonon 
polaritons mix phonons and photons \cite{Agranovich1982}. 

When the interfaces of these materials are bounded, the surface polaritons 
become localized. Localized surface phonon polaritons come along with 
surface optical phonons which were first studied with dielectric continuum 
models \cite{Englman1966,Fuchs1975}. Later on, with the progress made in nanofabrication 
techniques, surface optical phonons were studied due to their influence on 
the electron-phonon coupling in layered nanostructures  \cite{Licari1977,Wendler1985,Trallero1992,Nash1992,Comas1997}. 
Mutschkel et al. 
analyzed infrared properties of SiC particles of various polytypes 
observing size and shape-dependent resonances analogous to plasmon 
polaritons \cite{Mutschke1999}. Like plasmons, phonon polaritons can be used for applications 
regarding sensing \cite{Hillenbrand2002}, near-field optics \cite{Taubner2004} or superlensing \cite{Taubner2006}, 
but also  for thermal coherent infrared emission \cite{Greffet2002} and enhanced energy 
transfer \cite{Shen2009}. Moreover, in the last few years there is a great interest in phonon 
polaritons due to their intrinsic low losses in comparison with plasmon 
polaritons with many experiments probing phonon polariton properties of 
micro- and nano-patterned arrays of polar semiconductors \cite{Chen2014,Caldwell2015,Feng2015,Gubbin2016}.

Polaritons can be successfully described with dielectric continuum models by 
integrating Maxwell equations or Laplace equation in the quasi-static 
approximation \cite{Tsukerman2008}. Laplace equation suffices to describe polaritons if the 
size of the particle is much smaller than the wavelength of the incident 
light such that the electric field associated with the incident light is 
spatially uniform and retardation effects can be neglected. In fact, the 
first treatments of surface phonons and localized surface phonon polariton 
resonances were quasi-static \cite{Englman1966,Fuchs1975} which, as it will be seen in the following, 
are simpler by providing general 
properties for arbitrary shaped particles \cite{Fuchs1975}. 

Based on potential theory \cite{Kellogg1929,Khavison2007}, a boundary integral equation 
(BIE) method was developed in Ref. \cite{Fuchs1975}. The same theory was invoked 
independently for plasmon resonances \cite{Ouyang1989} and it reached wide 
recognition when it was used for localized plasmon resonances in 
metallic nanoparticles  \cite{Fredkin2003}. Needless to say, the formalism was applied also 
to radiofrequency properties of living cells \cite{Vrinceanu1996,Sandu2010}. Even though in most 
applications only a purely numerical treatment is possible, the BIE method 
provides direct information about the modes (the resonance strength and 
frequency) and their relations with particle shape and dielectric 
permittivity  \cite{Fuchs1975,Ouyang1989,Fredkin2003,Vrinceanu1996,Sandu2010}. 

In Ref. \cite{Englman1966} as well as in the following paper  \cite{Englman1968} Englman and Ruppin used the 
separation of variables method to solve the Laplace equation for spherical 
and cylindrical shapes together with their corresponding shells as well as 
the slab geometry as a limiting case of cylindrical shell. The BIE method in 
a separation of variables scheme was used for quantum wires of elliptical 
and circular sections  \cite{Knipp1992a} and quantum dots of spheroidal shape \cite{Knipp1992b}. Later 
on, the Laplace equation was solved with the separation of variables method 
for spheroidal \cite{Comas2002} and ellipsoidal \cite{Reese2004} shapes. 

The BIE method is associated with a non-symmetric boundary integral operator 
that is called the electrostatic operator, while its adjoint is called the 
Neumann-Poincar\'{e} operator \cite{Khavison2007}. The electrostatic operator is well 
behaved in the sense that if the boundary of the particle/domain is smooth, 
the operator is compact and its spectrum has discrete real eigenvalues 
\cite{Kellogg1929,Khavison2007}. This operator can be made self-adjoint using Plemelj 
symmetrization principle which, from practical point of view, relates the 
eigenfunctions of the electrostatic operator with those of its adjoint \cite{Sandu2013}. 
In the BIE method the eigenvalues of the electrostatic operator are related 
to plasmon resonances \cite{Fredkin2003}, with a straightforward eigenvalue-resonance 
relationship for Drude metals \cite{Sandu2013,Sandu2011}. Mathematical literature offers the 
whole set of eigenvalues and eigenfunctions for disks, ellipses, and spheres 
\cite{Khavison2007}, as well as for spheroids \cite{Ahner1986,Ahner1994a,Ahner1994b} and ellipsoids \cite{Ritter1995}. 

In this paper, similar to localized surface plasmon resonances (LSPRs) in 
metallic nanoparticles \cite{Sandu2013,Sandu2011} we establish explicit relationships between 
the localized surface phonon resonances (LSPhRs) and the eigenvalues of the 
electrostatic operator and dielectric properties of nanocrystals. 

While 
mathematical literature offer analytic forms of the eigenvalues and 
eigenfunctions for various shapes we will show a way of obtaining the spectral properties of 
the electrostatic operator from direct resolution of Laplace equation 
performed on specific physical problems regarding surface phonon or plasmon 
polaritons. Combining the BIE method with the separation of variables we 
recover all the eigenvalues embodied in the energies of surface phonons 
given in Refs. \cite{Englman1966,Englman1968,Knipp1992a,Knipp1992b,Comas2002,Reese2004} and in plasmon resonance frequencies found in Refs. \cite{Moussiaux1977,Prodan2004,Brandl2007,Morandi2008,Wan2013}. 
Thus, we unify all scattered results from 
literature and we put them in a general setting that can be used to 
interpret easily and intuitively the experiments concerning localized 
surface phonon polaritons. 

Additionally, the symmetry analysis of Laplace 
equation \cite{Boyer1976,Miller1977} has been rarely if ever exploited explicitly to characterize 
the electrostatic operator and its adjoint which are key elements of the 
resolution of Laplace equation. Essentially, we use the same separation of 
variables method to expand the free-space Green's function (also called the 
fundamental solution in mathematical literature) in an appropriate 
eigenfunction system. From Lie group-theoretical analysis it is known that the solutions 
of Laplace equation for each separable coordinate system 
are the eigenfunctions of a pair of two commuting operators \cite{Boyer1976,Miller1977,Makarov1967}. It 
turns out that these functions are also the eigenfunctions of the adjoint of 
the electrostatic operator and are proportional to the eigenfunctions of 
electrostatic operator itself. Therefore, we can also recover the 
eigenvectors for spheres, prolate and oblate spheroids, and ellipsoids. In 
addition to that we calculate both the eigenvalues and eigenfunctions for 
circular and elliptic cylinders. 

We further analyze the interaction of IR 
light with an array of GaN micro-disks studied in Ref. \cite{Feng2015} and we show that 
the results of the BIE method may offer more information than the calculations 
obtained with finite element methods. With the BIE method we also provide another 
interpretation of the scaling law of the plasmon resonance frequency with 
the size of graphene nano-disks \cite{Fang2013}. 

Finally, there is a new interest in 
spectral properties of the electrostatic operator regarding some specific 
applications like cloaking by anomalous localized resonance \cite{Nicorovici1994,Milton2005,Milton2006,Ando2016}, inverse 
problems \cite{Ammari2014,Ammari2016}, and materials characterization \cite{Ammari2013}. We will discuss how 
our results fit into these new areas of interest and how symmetry-based 
analysis may be expanded from electrostatic regime to the full-wave 
treatment of polaritons with the possibility of studying localized states of light in the continuum \cite{Silveirinha2014,Monticone2014,Hsu2016}.

The paper is structured as follows. In the next section we show the way by which the 
surface phonon and plasmon polariton resonances are calculated from the eigenvalues of the 
electrostatic operator. In section 3 we present two procedures of calculating the eigenvalues and 
the eigenfunctions of the electrostatic operator for shapes generated by 
separable coordinate systems. One procedure relies on the solutions of Laplace equation in 
separable coordinate systems and the other is based on the symmetry properties of Laplace equation 
applied to the electrostatic operator. In section 4 we discuss our results in the 
context of current research interests and future developments, and in section 
5 we conclude our work. 

\section{Surface plasmon versus surface phonon polariton resonances}

\begin{figure}
\centering
\includegraphics [width=6in,height=4in] {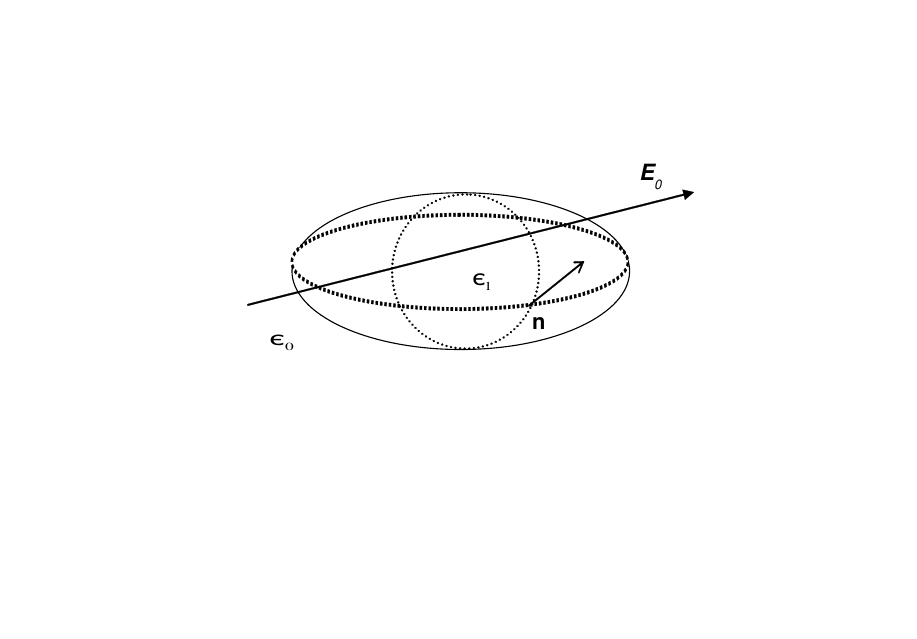} 
\caption{\label{fig:1}
A nanoparticle of permittivity $\epsilon_1 $ bounded by surface $\Sigma $ 
and surounded by a dielectric of permittivity $\epsilon_0 $. Here \textbf{n} is the normal to $\Sigma $ and $\bf{E}_0 $ is the applied field.}
\end{figure}

Continuum models describe successfully both plasmon and phonons polaritons. 
They are based on the same boundary conditions expressing the behaviour of 
optical modes at the interfaces. The boundary conditions are the continuity 
of the electric potential and of the normal component of electric induction. 
The model is schematically shown in Fig.~\ref{fig:1}, where a dielectric domain of 
volume $V$ and relative complex permittivity $\epsilon_1$ is bounded by surface \textit{$\Sigma $} and is 
surrounded by a medium of  relative complex dielectric permittivity $\epsilon_0 $. In the 
quasi-static limit, the applied field is uniform of strength $\bf{E}_0 $ and its 
time-dependence is assumed harmonic, i. e., a complex factor $\exp \left( { 
- i\omega t} \right)$ is understood.

In the local approximation the response of the system is given by a surface 
polarization charge density inducing a nanoparticle polarizability 
which has an eigenmode decomposition given by \cite{Sandu2011,Sandu2013}

\begin{equation}
\label{eq1}
\alpha = V\sum\limits_k {\frac{w_k }{\frac{1} {2\lambda } - \chi _k}} ,
\end{equation}

\noindent
where $\chi _k $ are the eigenvalues of the electrostatic operator

\begin{equation}
\label{eq2}
\hat {M}\left[ u \right] = - \int\limits_{\bf{x}' \in \Sigma } 
{u\left( {\bf{x}'} \right)\frac{\partial }{\partial \bf{n}_x }G\left( {\bf{x},\bf{x}'} 
\right)d\Sigma _{\bf{x}'} } 
\end{equation}

\noindent
while $w_k$ are subunitary numbers determined by $\bf{E}_0 $ and the eigenfunctions of 
$\hat {M}$ and its adjoint, and $\lambda = (\epsilon_1 - \epsilon_0)/(\epsilon_1 + \epsilon_0)$. In Eq. (\ref{eq2})

\begin{equation}
\label{eq3}
G\left( {\bf{x},\bf{x}'} \right) = \frac{1}{4\pi }\frac{1}{\left| {\bf{x} - \bf{x}'} \right|}
\end{equation}

\noindent
is the free-space Green's function and $\partial/\partial\bf{n}_x$ is the normal derivative. The electrostatic operator and its adjoint are defined on the Hilbert space $L^2(\Sigma)$ of square-integrable functions on $\Sigma $. The extinction cross-section is related to 
nanoparticle polarizability by \cite{Maier2007}

\begin{equation}
\label{eq4}
C_{ext} = \frac{2\pi }{\lambda }Im\left( \alpha \right).
\end{equation}

\noindent
Also, the near-field enhancement has an eigenmode decomposition similar to Eq. (\ref{eq1}) \cite{Sandu2013}. 

If we assume that the dielectric permittivity $\epsilon_0$ of the embedding medium is a real constant $\varepsilon _{d} $ and the metallic nanoparticle is a Drude metal whose permittivity is

\begin{equation}
\label{eq5}
\epsilon_1 = \varepsilon_{m} - \frac{\omega _{p}^2 }{\omega \left( 
{\omega + \mbox{i}\gamma } \right)},
\end{equation}

\noindent
we can obtain compact relations between the polarizability $\alpha _{LSPR} 
$ characterizing the localized plasmon resonances (LSPRs) and the 
eigenvalues of $\hat {M}$ \cite{Sandu2013} 

\begin{widetext}
\begin{equation}
\label{eq6}
\alpha_{LSPR} (\omega) = 
V\sum_k \frac{w_k (\varepsilon_m - \varepsilon_{d})}{\varepsilon_{k}}-\frac{w_k}{1/2-\chi_k} \frac{\varepsilon_{d}}{\varepsilon_{k}}
\frac{\tilde\omega_{pk}^2}{\omega(\omega + i\gamma) - \tilde\omega_{pk}^2}.
\end{equation}
\end{widetext}
Here $\varepsilon _{k} = (1/2 + \chi _k) \varepsilon_{d} + (1/2 - \chi_k) \varepsilon _m$ is an 
effective dielectric constant and

\begin{equation}
\label{eq7}
\tilde {\omega }_{pk} = \omega _p \sqrt {{\left( {1 / 2 - \chi _k } \right)} 
\mathord{\left/ {\vphantom {{\left( {1 / 2 - \chi _k } \right)} {\varepsilon 
_{k} }}} \right. \kern-\nulldelimiterspace} {\varepsilon _{k} }} 
\end{equation}

\noindent
is the expression of LSPR frequency. Equations (\ref{eq6})-(\ref{eq7}) show clearly the 
relation between localized surface plasmon resonances and the eigenvalues of 
$\hat {M}$. For instance, if $\varepsilon _m = \varepsilon _{d} = 1$ then 
$\tilde {\omega }_{pk} = \omega _p \sqrt {1 / 2 - \chi _k } $. On the other 
hand, the resonance strength is included in the numerical factor $w_k $. 

A similar expression can be obtained for a void in which $\epsilon_0$ takes a 
Drude form (\ref{eq4}) and $\epsilon_1 $ is a real constant. The expressions for voids are 
obtained by the swap $\left( {1 \mathord{\left/ {\vphantom {1 2}} \right. 
\kern-\nulldelimiterspace} 2 - \chi _k } \right) \leftrightarrow \left( {1 
\mathord{\left/ {\vphantom {1 2}} \right. \kern-\nulldelimiterspace} 2 + 
\chi _k } \right)$ and replacing $w_k $ with $ - w_k $. Often in numerical 
simulation, instead of a Drude metal, experimental values are considered for 
metals like gold and silver \cite{Maier2007}. The main difference between experimental 
values and the Drude model of dielectric permittivity is the interband 
contributions that become significant for wavelengths below 500 nm. 

In polar crystals the frequency band between the frequency of transverse 
optical phonons ($\omega _T )$ and the frequency of transverse optical 
phonons ($\omega _L )$ is called the reststrahlen band, where the dielectric 
permittivity is negative and can be modeled as a Lorentz oscillator \cite{Kittel1996}

\begin{equation}
\label{eq8}
\epsilon = \varepsilon _\infty \left( {1 + \frac{\omega _{L}^2 - \omega 
_{T}^2 }{\omega _{T}^2 - \omega ^2 + {i}\Gamma \omega }} 
\right).
\end{equation}
In the reststrahlen band localized surface phonon polaritons of polar 
nanocrystals have similar behaviour as localized surface plasmons in 
metallic nanoparticles. Similar to Eqs. (\ref{eq6})-(\ref{eq7}) we can derive the 
phonon polarizability that defines the LSPhRs. The phonon polarizability 
is obtained by the 
following algebraic substitutions: $\omega ^2 \to \omega ^2 - \omega _{T}^2 $, 
$\omega _{p}^2 \to \varepsilon _\infty \left( {\omega _{L}^2 - 
\omega _{T}^2 } \right)$, $\varepsilon _{m} \to \varepsilon 
_\infty $, and $i \gamma \omega \to -i \Gamma \omega$. Thereby, the counterpart of Eq. (\ref{eq6}), which is the polarizability $\alpha_{LSPhR} $ characterizing the 
LSPhRs, takes the form 

\begin{widetext}
\begin{equation}
\label{eq9}
\alpha_{LSPhR} (\omega) = 
V\sum_k \frac{w_k (\varepsilon_\infty - \varepsilon_{d})}{\varepsilon_{k}}-\frac{w_k}{1/2-\chi_k} \frac{\varepsilon_{d}}{\varepsilon_{k}}
\frac{\tilde\omega_{phk}^2}{\omega(\omega - i\Gamma) - \omega _{T}^2 - \tilde\omega_{phk}^2},
\end{equation}
\end{widetext}

\noindent
where 

\begin{equation}
\label{eq10}
\tilde {\omega }_{phk} = \sqrt {{\varepsilon _\infty \left( {1 / 2 - \chi _k 
} \right)\left( {\omega _{L}^2 - \omega _{T}^2 } \right)}/ {\varepsilon _{k} }}
\end{equation}

\noindent
is the LSPhR frequency and  $\varepsilon _{k} = (1/2 - \chi _k) \varepsilon _\infty + (1/2 + \chi _k) \varepsilon _{d}$. Equation (\ref{eq10}) illustrates the dependence of LSPhRs on the 
nanocrystal shape in the same fashion as Eq (\ref{eq7}) shows the dependence of 
LSPR on the metallic nanoparticle shape. For a nanovoid in a polar crystal 
$\varepsilon _0 $ is replaced by Eq. (\ref{eq8}), $\varepsilon _1$ is a real 
constant denoted by $\varepsilon _{d} $, and 

\begin{widetext}
\begin{equation}
\label{eq11}
\alpha_{LSPhR\_v} (\omega) = 
V\sum_k \frac{w_k (\varepsilon_{d} - \varepsilon_\infty)}{\varepsilon_{k\_v}} + \frac{w_k}{1/2+\chi_k} \frac{\varepsilon_{d}}{\varepsilon_{k\_v}}
\frac{\tilde\omega_{phk\_v}^2}{\omega(\omega - i\Gamma) - \omega _{T}^2 - \tilde\omega_{phk\_v}^2},
\end{equation}
\end{widetext}

\noindent
with $\tilde {\omega }_{phk\_v} = \sqrt {{\varepsilon _\infty \left( {1 / 2 + \chi _k 
} \right)\left( {\omega _{L}^2 - \omega _{T}^2 } \right)}/ {\varepsilon _{k\_v} }}$ 
and $\varepsilon _{k\_v} = (1/2 - \chi _k) \varepsilon _{d} + (1/2 + \chi _k) \varepsilon _\infty$. 
We have given explicitly the expressios of LSPhRs for both nanocrystals and 
nanovoids because often on can encounter both situations like semiconductor 
nanodots embedded in a crystal host that has its own reststrahlen band. To 
conclude this section, with the boundary integral equation method we may 
calculate the eigenvalues and eigenvectors of the electrostatic operator to 
obtain a set of numbers $\left\{ {\chi _k ,w_k } \right\}$ which provides 
the localized surface plasmon/phonon polariton spectrum with the help of Eqs. (\ref{eq6})-(\ref{eq11}).

\section{Eigenvalues and eigenfunctions of the electrostatic operator by separation of variables method}
\subsection{Retrieval of eigenvalues from the solutions of Laplace equation}

As we have already mentioned the whole set of eigenvalues and eigenfunctions 
of the electrostatic operator is known for disks, ellipses (see also \cite{Ahner1986b} 
for an earlier account), spheres \cite{Khavison2007}, spheroids \cite{Ahner1986,Ahner1994a,Ahner1994b} 
and ellipsoids \cite{Ritter1995}. 
These calculations are based on the expansion of the free-space Green's 
function in a separated form of the corresponding coordinate systems i. e., 
spherical, spheroidal, and ellipsoidal coordinate systems \cite{Moon1961}. The separation of variables problem dates 
back to the end of 19th century \cite{Stackel1891, Bocher1894} when the BIE method was laid out on solid ground 
(a historic account can be found in \cite{Costabel2007}) with important contributions made by Robertson 
\cite{Robertson1927} and Eisenhart \cite{Eisenhart1934a,Eisenhart1934b}. The scalar Helmholtz equation and subsequently 
the Laplace equation are separable in eleven coordinate systems, while the Laplace equation is $R$-separable in 
other six coordinate systems (see \cite{Boyer1976,Miller1977} for a detailed 
discussion on separability and $R$-separability of Laplace equation) These 
coordinates are called cyclidic coordinates which are quartic 
surfaces \cite{Moon1961,Miller1977}. In Refs. \cite{Englman1968,Knipp1992a,Knipp1992b,Comas2002,Reese2004} 
the separation of variables method 
is invoked to solve the Laplace equation associated with a specific 
physical problem. We consider a separable coordinate system is given by 
coordinates $\left( {\xi _1 ,\xi _2 ,\xi _3 } \right)$ and the surface 
\textit{$\Sigma $} is defined by the equation $\xi _3 = \xi _0 $, where $\xi _0 $ is a 
constant. Different values of $\xi _3 $ generate confocal surfaces. The 
boundary conditions are determined by the continuity of the normal component 
of electric induction \textbf{\textit{D}} at the surface $\xi _3 = \xi _0 $ 
which reads  \cite{Knipp1992a,Knipp1992b,Comas2002,Reese2004}

\begin{equation}
\label{eq12}
\frac{\epsilon_1 \left( \omega \right)}{\epsilon_0 } = f_{lm} \left( {\xi _0 } \right).
\end{equation}

\noindent
Here $f_{lm} \left( {\xi _0 } \right)$ is the ratio of logarithmic 
derivatives of two functions, $P_{lm} \left( {\xi _3 } \right)$ and 
$Q_{lm} \left( {\xi _3 } \right)$. The indices $l$ and $m$ define these functions which are factors of Laplace 
equation solutions in the separable coordinate system \cite{Knipp1992a,Knipp1992b,Comas2002,Reese2004}. We analyze 
Eq. (\ref{eq12}) from BIE point of view. A close inspection of Laplace equation in separable coordinate 
systems shows that, for each $l$ and $m$, the solution of Laplace equation is 
also proportional to the eigenfunction $u_{lm} ({\xi _1 ,\xi _2 })$ of $\hat {M}$ such that the continuity on the normal component of 
\textbf{\textit{D}} reduces to 

\begin{equation}
\label{eq13}
\frac{\epsilon_1 \left( \omega \right)}{\epsilon_0 } = - \frac{1/2 + \chi _{lm} }{1/2 - 
\chi _{lm} },
\end{equation}

\noindent
where $\chi _{lm}$ is the eigenvalue of $\hat {M}$ corresponding to $u_{lm} 
\left( {\xi _1 ,\xi _2 } \right)$. Relation (\ref{eq13}) has a general validity for 
arbitrary shapes and it may be used not only for surface phonon polaritons 
but also for surface plasmons. Combining (\ref{eq8}) and (\ref{eq13}) we may obtain the 
resonance frequency of surface phonon polaritons not only when the surface 
is described by a separable coordinate system but also in general for 
arbitrary shape

\begin{widetext}
\begin{equation}
\label{eq14}
\tilde {\omega }_{lm}^2 = \omega _T^2 \frac{\left( {1 
\mathord{\left/ {\vphantom {1 2}} \right. \kern-\nulldelimiterspace} 2 + 
\chi _{lm} } \right)\varepsilon _{d} + \left( {1 \mathord{\left/ {\vphantom 
{1 2}} \right. \kern-\nulldelimiterspace} 2 - \chi _{lm} } 
\right)\varepsilon _0 }{\left( {1 \mathord{\left/ {\vphantom {1 2}} \right. 
\kern-\nulldelimiterspace} 2 + \chi _{lm} } \right)\varepsilon _{d} + 
\left( {1 \mathord{\left/ {\vphantom {1 2}} \right. 
\kern-\nulldelimiterspace} 2 - \chi _{lm} } \right)\varepsilon _\infty }.
\end{equation}
\end{widetext}

\noindent
Here $\varepsilon _0 $ is the static dielectric constant of the crystal 
which is linked to $\varepsilon _\infty $ by Lyddane-Sachs-Teller relation 
\cite{Kittel1996}

\begin{equation}
\label{eq15}
\frac{\varepsilon _0 }{\varepsilon _\infty } = \frac{\omega _L^2 }{\omega 
_T^2 }.
\end{equation}

Equation (\ref{eq14}) is another form of Eq. (\ref{eq10}) via Lyddane-Sachs-Teller relation. From (\ref{eq14}) it is easy to check that if there are shapes with $\left( {1 
\mathord{\left/ {\vphantom {1 2}} \right. \kern-\nulldelimiterspace} 2 - 
\chi _{lm} } \right) \to 0$ then $\tilde {\omega }_{lm} \to \omega _T $. We 
will see in the next section that infinitely long nanorods fulfill this 
condition. In the same time, infinitely thin nanodisks have an eigevalue 
fulfilling the above condition and other satisfying the condition $\left( {1 
\mathord{\left/ {\vphantom {1 2}} \right. \kern-\nulldelimiterspace} 2 + 
\chi _{lm} } \right) \to 0$ with the corresponding resonance frequency 
$\tilde {\omega }_{lm} \to \omega _L $.

\subsection{The eigenvalues and the eigenfunctions of the electrostatic operator from symmetry analysis}

In this subsection we will present a symmetry-based procedure by which the eigenvalues and the eigenfunctions 
of the electrostatic operator are calculated for various shapes originating from separable coordinate systems 
admitting. According to the program linking symmetry with 
separation of variables the solutions for each coordinate system are common 
eigenfunctions of a pair of commuting operators \cite{Miller1977}. It turns out that the 
eigenfunctions of the electrostatic operator are generated by the 
eigenfunctions of these two commuting operators. This fact can be seen by 
expressing the free-space Green's function (\ref{eq3}) in a separated form, which is 
determined by three second-order differential equations and two separation 
constants. The eigenvalues of the electrostatic operator are given by the 
third differential equation in the third coordinate $\xi _3 $. The third 
differential equation, an ordinary second order equation, has two solutions 
of which one is regular in origin (inside the surface) and the other is 
regular at infinity or outside the surface. The two solutions for $\xi _3 
= \xi _0 $ will provide the eigenvalues. This recipe is applied in the 
following. We designate by $l,$ and $m$ the indices defining the eigenvalues of 
the two commuting operators. These indices also define the eigenvalues of 
the electrostatic operator. Let $P_{lm} \left( {\xi _3 } \right)$ and 
$Q_{lm} \left( {\xi _3 } \right)$ be, respectively, the regular solution in 
origin and at infinity for the third differential equation. According to 
\cite{Knipp1992a,Knipp1992b,Comas2002,Reese2004} $f_{lm} \left( {\xi _0 } \right)$ is the ratio of logarithmic 
derivatives of $P_{lm} \left( {\xi _3 } \right)$ and $Q_{lm} \left( {\xi _3 
} \right)$. Then with the help of (\ref{eq12}) and (\ref{eq13}) we may easily obtain the 
eigenvalues as

\begin{equation}
\label{eq16}
\chi _{lm} = \frac{1}{2} + \frac{P_{lm} '\left( {\xi _3 } \right)_{\xi _3 = 
\xi _0 } Q_{lm} \left( {\xi _3 } \right)_{\xi _3 = \xi _0 } }{W\left[ 
{P_{lm} \left( {\xi _3 } \right),Q_{lm} \left( {\xi _3 } \right)} 
\right]_{\xi _3 = \xi _0 } },
\end{equation}

\noindent
where the prime means the derivative and $W\left \{ {P_{lm} ,Q_{lm} } \right \} = P_{lm} Q_{lm}' - P_{lm}' Q_{lm} $ 
is the wronskian \cite{Morse1953} of $P_{lm} \left( {\xi _3 } \right)$ and $Q_{lm} 
\left( {\xi _3 } \right)$, all evaluated at $\xi _3 = \xi _0 $. Hence direct 
resolution of Laplace equation performed in \cite{Knipp1992a,Knipp1992b,Comas2002,Reese2004} leads to the full set of 
the eigenvalues of $\hat {M}$ for the shapes considered in those papers.

The explicit expressions of the eigenvalues and the eigenfunctions are given in 
mathematical literature for spheres, spheroids and ellipsoids  \cite{Ahner1986,Ahner1994a,Ahner1994b,Ritter1995}. These 
calculations are based on the expression of the Green's function (\ref{eq3}) in the 
separable coordinate systems. A similar approach was used for disks and 
ellipses in two-dimensional (2D) space \cite{Ahner1986b}. Below we will also provide explicit 
expressions for the eigenvalues and the eigenfunctions of the electrostatic 
operator for three-dimensional (3D) counterparts of disks and ellipses, i. e., 
circular and elliptic cylinders, which to our knowledge is new. In 
principles, we have to solve the equation

\begin{equation}
\label{eq17}
\Delta G\left( {\bf{x},\bf{x}'} \right) = - \delta \left( {\bf{x} - \bf{x}'} \right)
\end{equation}

\noindent
in an appropriate coordinate systems by separation of variables method. The 
caveat of our analysis and calculations is the fact that the eigenfunctions of the two 
commuting operators associated with each separable coordinate system form a 
basis in which the free-space Green's function may be expressed and 
consequently the electrostatic operator can become diagonal. In fact the 
adjoint operator is diagonal in this basis, while the eigenfunction of the 
electrostatic operator are only proportional to these functions. 
This is not a surprise since the adjoint of the electrostatic operator is used in potential theory 
to generate the solution of Laplace equation with Dirichlet boundary conditions. 
In the previous subsection we have effectively retrieved the eigenvalues of the electrostatic 
operator from the solution of Laplace equation for surface phonons 
and plasmons. 
In the 
following we present also a systematic way to 
obtain not only the eigenvalues but also the eigenfunctions of the electrostatic operator and its 
adjoint for shapes like spheres, prolate and oblate 
spheroids, ellipsoids, and elliptic and circular cylinders. 

(a) sphere 

Working in spherical coordinates, $x = \rho \sin \theta \cos 
\varphi ,\;y = \rho \sin \theta \sin \varphi ,\;z = \rho \cos \theta $, the 
sphere is evidently determined by $\rho = \rho _0 $, while the 
eigenfunctions of the electrostatic operator are spherical harmonics $Y_{lm} 
\left( {\theta ,\varphi } \right)$ and the eigenvalues are

\begin{equation}
\label{eq18}
\chi _{lm} = \frac{1}{2}\frac{1}{2l + 1}.
\end{equation}
Here $l$ is a natural number and $m$ is integer with $\left| m \right| \le l$. 
In 3D sphere is the only surface for which the electrostatic operator is 
symmetric \cite{Khavison2007}. Expression (\ref{eq18}) can be also deduced using Eq. (\ref{eq13}) from Ref. 
 \cite{Comas2002}.

(b) prolate spheroid 

In these coordinate system the transformations are: 
$x = a\sqrt {({\eta ^2 - 1})({1 - \zeta ^2})}\cos\varphi$, 
$y = a\sqrt {({\eta ^2 - 1})({1 - \zeta ^2})} \sin \varphi$, $z = a\eta \zeta $ 
with $\eta \ge 1$ and $ - 1 \le \zeta \le 1$. Prolate spheroids characterized by equation $\eta = \eta _0 $ 
have the eigenfunctions of the electrostatic operator proportional to 
$e^{im\varphi }P_{lm} \left( \zeta \right)$ and its eigenvalues equal to

\begin{equation}
\label{eq19}
\chi _{lm} = \frac{1}{2} - (-1)^m \frac{\left( {l - m} \right)!}{\left( {l + m} 
\right)!}\left( {\eta _0^2 - 1} \right)P_{lm} \left( {\eta _0 } 
\right)'Q_{lm} \left( {\eta _0 } \right)
\end{equation}

\noindent
where $P_{lm} $ and $Q_{lm} $ are the associated Legendre functions, $l$ is a 
natural number and $m$ is integer with $\left| m \right| \le l$. We considered 
the definitions of $P_{lm} $ and $Q_{lm} $ such that their wronskian is 
$W\left\{ {P_{lm} \left( \eta \right),Q_{lm} \left( \eta \right)} \right\} = 
(-1)^m {\left( {l + m} \right)!} \mathord{\left/ {\vphantom {{\left( {l + m} 
\right)!} {\left( {\left( {l - m} \right)!\left( {1 - \eta ^2} \right)} 
\right)}}} \right. \kern-\nulldelimiterspace} {\left( {\left( {l - m} 
\right)!\left( {1 - \eta ^2} \right)} \right)}$ \cite{Lebedev1965}.

(c) oblate spheroid 

The oblate spheroidal coordinates are obtained by 
replacing $a $ with \textit{--ia} and $\eta $ with $i\eta $. Hence the eigenfunctions of the 
electrostatic operator are proportional to $e^{im\varphi }P_{lm} \left( 
\zeta \right)$ and the eigenvalues are

\begin{equation}
\label{eq20}
\chi _{lm} = \frac{1}{2} + (-1)^m \frac{\left( {l - m} \right)!}{\left( {l + m} 
\right)!}\left( {\eta _0^2 + 1} \right)P_{lm} \left( {i\eta _0 } 
\right)'Q_{lm} \left( {i\eta _0 } \right).
\end{equation}
The eigenvalues of $\hat {M}$ for prolate and oblate spheroids as well as its 
eigenfunctions are calculated in  \cite{Ahner1986,Ahner1994a,Ahner1994b}. These eigenvalues can also be 
retrieved with Eq. (\ref{eq13}) from Refs. \cite{Knipp1992b,Comas2002}.

(d) ellipsoid 

The ellipsoidal coordinates, appropriate for an ellipsoid with axes $a > b 
> c$, are the solutions \textit{$\theta $} of the equation

\begin{equation}
\label{eq21}
\frac{x^2}{a^2 + \theta } + \frac{y^2}{b^2 + \theta } + \frac{z^2}{c^2 + 
\theta } = 1,
\end{equation}

\noindent
which are denoted as $\rho \in \left[ {\left. { - c^2,\infty } \right)} 
\right.$, $\mu \in \left[ { - b^2, - c^2} \right]$, and $\nu \in \left[ { - 
a^2, - b^2} \right]$. The Cartesian coordinates are then expressed in 
ellipsoidal coordinates by the following

\begin{equation}
\label{eq22}
\begin{array}{l}
 x^2 = \frac{\left( {a^2 + \rho } \right)\left( {a^2 + \mu } \right)\left( 
{a^2 + \nu } \right)}{\left( {a^2 - b^2} \right)\left( {a^2 - c^2} \right)} 
\\ 
 y^2 = \frac{\left( {b^2 + \rho } \right)\left( {b^2 + \mu } \right)\left( 
{b^2 + \nu } \right)}{\left( {b^2 - c^2} \right)\left( {b^2 - a^2} \right)} 
\\ 
 z^2 = \frac{\left( {c^2 + \rho } \right)\left( {c^2 + \mu } \right)\left( 
{c^2 + \nu } \right)}{\left( {c^2 - a^2} \right)\left( {c^2 - b^2} \right)}. 
\\ 
 \end{array}
\end{equation}
The ellipsoid is governed by simple equation $\rho = \rho _0 $. Laplace 
equation separates in this coordinate system such that for each ellipsoidal 
coordinate a Lam\'{e} equation \cite{Arscott1962} of the form

\begin{equation}
\label{eq23}
4\Delta \left( \rho \right)\frac{d}{d\rho }\left\{ {\Delta \left( \rho 
\right)\frac{dL\left( \rho \right)}{d\rho }} \right\} = \left( {n\left( {n + 
1} \right) + B} \right)L\left( \rho \right)
\end{equation}

\noindent
is obeyed by all three ellipsoidal coordinates. Here $n$ is a natural number, $B$, 
which is a separation constant, is real \cite{Ritter1995,Dobner1998}, and $\Delta(\rho) = \sqrt{(\rho +a^2)(\rho + b^2)(\rho + c^2)}$. 
For each $n$ we obtain $2n + 1$ values of $B$ denoted as $B_n^m $ with $m = 0,1,\ldots \ldots ,2n$. Thus, 
the solution  $L_n^m $ is the first kind Lam\'{e} function of order $n$. 
This solution is regular, the second solution, which is regular at infinity, 
is obtained by standard procedure \cite{Morse1953}

\begin{equation}
\label{eq24}
K_n^m \left( \rho \right) = \left( {2n + 1} \right)L_n^m \left( \rho 
\right)\int\limits_\rho ^\infty {\frac{d\rho '}{\left[ {L_n^m \left( {\rho 
'} \right)} \right]^2\Delta \left( {\rho '} \right)}}. 
\end{equation}
The Wronskian considered here reads $W\left\{ {L_n^m \left( \rho 
\right),K_n^m \left( \rho \right)} \right\} = -{\left( {2n + 1} \right)} 
\mathord{\left/ {\vphantom {{\left( {2n + 1} \right)} {\Delta \left( \rho 
\right)}}} \right. \kern-\nulldelimiterspace} {\Delta \left( \rho \right)}$. 
All the above lead to the following eigenvalues

\begin{equation}
\label{eq25}
\chi _{nm} = \frac{1}{2} - \frac{\Delta \left( \rho _0 \right)L_n^m \left( 
{\rho _0 } \right)' K_n^m \left( {\rho _0 } \right)}{2n + 1}.
\end{equation}

The eigenfunctions of the electrostatic operator are proportional to $L_n^m 
\left( \mu \right)L_n^m \left( \nu \right)$. We have adopted the definition 
of ellipsoidal coordinates used in \cite{Ritter1995} and in the standard textbooks of 
electrodynamics like \cite{Stratton1941}. It provides the straightforward calculation of 
the wronskian and of the second kind Lam\'{e} function. Slightly different 
definitions provided in \cite{Moon1961} and used in \cite{Reese2004} give 
the same results \cite{Feng2016}. 
The eigenvalues (\ref{eq25}) 
were first calculated in \cite{Ritter1995} and can be deduced with the help of (\ref{eq13}) from 
Ref. \cite{Reese2004}. 
\noindent

(e) elliptic cylinder 

Such cylinders are suitably described in elliptic 
cylindrical coordinates: $x = d\cosh \left( u \right)\cos \left( v 
\right)$, $\;y = d\sinh \left( u \right)\sin \left( v \right)$, $\;z = z$, with 
$d$ an arbitrary positive constant. The surface of the cylinder is given by 
equation $u = u_0 $. For convenience we take $d = \sqrt {R^2 - r^2} $, where 
$R$ and $r$ are the lengths of the longer and, respectively, shorter semi-axis of 
the cross-sectional ellipsis. Accordingly, the constant $u_0 $ is equal to 
${\ln \left[ {{\left( {R + r} \right)} \mathord{\left/ {\vphantom {{\left( 
{R + r} \right)} {\left( {R - r} \right)}}} \right. 
\kern-\nulldelimiterspace} {\left( {R - r} \right)}} \right]} 
\mathord{\left/ {\vphantom {{\ln \left[ {{\left( {R + r} \right)} 
\mathord{\left/ {\vphantom {{\left( {R + r} \right)} {\left( {R - r} 
\right)}}} \right. \kern-\nulldelimiterspace} {\left( {R - r} \right)}} 
\right]} 2}} \right. \kern-\nulldelimiterspace} 2$. The eigenvalues of the 
electrostatic operator can be calculated with Eq. (\ref{eq13}) from Ref. \cite{Knipp1992a}. Also 
we will see below the separation of variables method provides the eigenvectors of 
the electrostatic operator which are proportional to either 
$e^{ikz}ce_m \left( {q,v} \right)$ or $e^{ikz}se_m \left( {q,v} \right)$. 
Here $k$ is real number and $ce_m \left( {q,v} \right)$, $se_m \left( {q,v} 
\right)$ are respectively the ``even'' and the ``odd'' solutions of the 
Matheiu equation \cite{Knipp1992a,McLachlan1947}

\begin{equation}
\label{eq26}
\frac{d^2V\left( {q,v} \right)}{dv^2} + \left( {\alpha - 2q\cos \left( {2v} 
\right)} \right)V\left( {q,v} \right) = 0,
\end{equation}

\noindent
where $q = - k^2\left( {R^2 - r^2} \right)$ and \textit{$\alpha $} is a separation constant 
which takes a characteristic value $a_m $ or $b_m $ of the $m^{th}$ solution 
$ce_m \left( {q,v} \right)$ or $se_m \left( {q,v} \right)$, respectively. If 
the characteristic values of the Mathieu equation are $a_m $ the eigenvalues 
are

\begin{equation}
\label{eq27}
\chi _{qm} = \frac{1}{2} + \frac{Ce_m \left( {q,u_0 } \right)'Fe_m \left( {q,u_0 
} \right)}{W_e(q)},
\end{equation}

\noindent
where $Ce_m \left( {q,u_0 } \right)$ and $Fe_m \left( {q,u_0 } \right)$ are 
the solutions of the corresponding associated Mathieu equation

\begin{equation}
\label{eq28}
\frac{d^2U\left( {q,u} \right)}{dv^2} - \left( {\alpha - 2q\cosh \left( {2u} 
\right)} \right)U\left( {q,u} \right) = 0.
\end{equation}

Here \textit{$\alpha $} takes the characteristic value $a_m $. Similarly, if \textit{$\alpha $} is the 
characteristic value $b_m $ the eigenvalues are

\begin{equation}
\label{eq29}
\chi _{qm} = \frac{1}{2} + \frac{Se_m \left( {q,u_0 } \right)'Ge_m \left( {q,u_0 
} \right)}{W_o(q)},
\end{equation}

\noindent
with $Se_m \left( {q,u_0 } \right)$ and $Ge_m \left( {q,u_0 } \right)$ the 
solution of (\ref{eq28}) when \textit{$\alpha $} takes the characteristic value $b_m $. We denoted by $W_e(q)$ 
the wronskian of $Ce_m(q,u)$ and $Fe_m(q,u)$ and by $W_o(q)$ the wronskian of $Se_m(q,u)$ and $Ge_m(q,u)$.
The wronskians of the solutions of both the Mathieu equation and the associated Mathieu equation are constant. On 
the other hand the Mathieu equation has a second solution which is non-periodic. We can take the second solution 
to be odd or even if the first (periodic) solution is even or odd \cite{ Arscott1964}. Thus, the wronskians of 
the solutions of Mathieu equation as well as of associated Mathieu equation depend only on $q$ 
(see sections 28.5.8, 28.5.9 and 28.20.3-28.20.7 of  \cite{ Olver2010}). The eigenvalues expressed by (\ref{eq27}) and 
(\ref{eq29}) are consistent with the surface phonon spectrum given in Ref. \cite{Knipp1992a}. 

The eigenvalues as well as the eigenfunctions of the electrostatic operator 
for elliptic cylinder can be conveniently calculated by expressing the 
free-space Green's function $G\left( {\bf{x},\bf{x}'} \right)$ in separable elliptic 
cylindrical coordinates. First, we use the identity $\frac{1}{2\pi 
}\int\limits_{ - \infty }^\infty {e^{ik\left( {z - z'} \right)}dk} = \delta 
\left( {z - z'} \right)$. Then, it can be shown that the Mathieu functions 
$\left\{ {ce_m \left( {q,v} \right),se_m \left( {q,v} \right)} \right\}$ are 
not only orthogonal on the space of square-integrable function $L^2\left( { 
- \pi ,\pi } \right)$, but also complete since they are the solutions of a 
Sturm-Liouville problem \cite{Higgins1977}. Hence the completeness relation on $L^2\left( 
{ - \pi ,\pi } \right)$ can take the form 
$\frac{1}{\pi} \sum\limits_{m = 0}^\infty {ce_m 
\left( {q,v} \right)ce_m \left( {q,v'} \right) + \sum\limits_{m = 1}^\infty 
{se_m \left( {q,v} \right)se_m \left( {q,v'} \right)}} = \delta \left( {v - 
v'} \right) $. 
Based on the separation of variables 
method we then may show that the free-space Green's function $G\left( {\bf{x},\bf{x}'} 
\right)$ depending on $\left( {u,v,z} \right)$ and $\left( {u',v',z'} 
\right)$ has the following expression in elliptic cylindrical 
coordinates:

\begin{eqnarray}\nonumber \label{eq30}
G = \frac{-1}{2\pi^2}\int\limits_{ - \infty }^{\infty} {dk\,e^{ik(z - z')} }  
\{\sum \limits_{m = 0}^{\infty} \frac{1}{W_e(q)}{ce_m (q,v)ce_m(q,v')Ce_m(q,u_< )Fe_m(q,u_>)} + \\ 
\sum \limits_{m = 1}^{\infty} \frac{1}{W_o(q)}{se_m(q,v)se_m(q,v')Se_m (q,u_<)Ge_m (q,u_>)}
\}. 
\end{eqnarray}

\noindent
Here $u_ < (u_ > )$ is the smaller (larger) of $u$ and $u'$, $k$ is real and $q = - 
k^2\left( {R^2 - r^2} \right)$. The normal derivative to elliptic 
cylindrical surface is

\begin{equation}
\label{eq31}
\frac{\partial }{\partial n} = \frac{1}{d\left( {\sinh \left( u \right)^2 + 
\sin \left( v \right)^2} \right)^{1 \mathord{\left/ {\vphantom {1 2}} 
\right. \kern-\nulldelimiterspace} 2}}\frac{\partial }{\partial u}.
\end{equation}

Then it is easy to check that the eigenvalues are those expressed by Eqs. 
(\ref{eq27}) and (\ref{eq29}) and the eigenfunctions of the $\hat {M}$ are

\begin{equation}
\label{eq32}
\frac{e^{ikz}}{\left( {\sinh \left( u \right)^2 + \sin \left( v \right)^2} 
\right)^{1 \mathord{\left/ {\vphantom {1 2}} \right. 
\kern-\nulldelimiterspace} 2}}\left\{ {{\begin{array}{*{20}c}
 {ce_m \left( {q,v} \right)} \hfill \\
 {se_m \left( {q,v} \right)} \hfill \\
\end{array} }} \right\},
\end{equation}

\noindent
while the eigenfunctions of its adjoint have the same expression without the 
factor $1 \mathord{\left/ {\vphantom {1 {\left( {\sinh \left( u \right)^2 + 
\sin \left( v \right)^2} \right)^{1 \mathord{\left/ {\vphantom {1 2}} 
\right. \kern-\nulldelimiterspace} 2}}}} \right. \kern-\nulldelimiterspace} 
{\left( {\sinh \left( u \right)^2 + \sin \left( v \right)^2} \right)^{1 
\mathord{\left/ {\vphantom {1 2}} \right. \kern-\nulldelimiterspace} 2}}$. 
These results are the 3D generalization of 2D ellipses \cite{Khavison2007,Ahner1986b,Ando2016} which 
can be recovered if we take $k = q = 0$. So, if we consider $k = q = 0$, then 
$a_m = b_m = m^2$, $ce_m \left( {q,v} \right) = \cos \left( {mv} \right)$, 
$se_m \left( {q,v} \right) = \sin \left( {mv} \right)$, and the eigenvalues 
are

\begin{equation}
\label{eq33}
\chi _m = \frac{1}{2e^{2m\,u_0 }}.
\end{equation}
During the preparation of the manuscript a recent paper has come to our 
attention \cite{Cohl2012}. In that paper the eigenfunction expansions of the 
free-space Green's function are calculated in elliptic and parabolic cylinder 
coordinates \cite{Cohl2012}. The authors used a different proof based on Lipschitz and 
Lipschitz-Hankel integral identities but their expansion given in theorem 
4.3 is similar to our expression (\ref{eq30}). Also, in theorem 5.3 the authors used other 
two linearly independent solutions of the associated Mathieu equations whose wronskian is -1.

(f) circular cylinder 

In cylindrical coordinates $x = \rho \sin \theta ,\;y 
= \rho \cos \theta ,\;z = z$ the cylinder with circular cross-section is 
described by $\rho = \rho _0 $. The eigenvalues and eigenfunctions of the 
electrostatic operator can be easily calculated if we express the free-space 
Green's function in cylindrical coordinates. The Green's function has a well 
known expression in cylindrical coordinate that is given, for example, in 
Jackson's textbook \cite{Jackson1975}. In cylindrical coordinates $\left( {\rho 
,\theta ,z} \right)$ and $\left( {\rho ',\theta ',z'} \right)$  the free-space 
Green's function has the 
following form 

\begin{widetext}
\begin{equation}
\label{eq34}
G = \frac{1}{4\pi ^2}\int\limits_{ - \infty }^\infty {dk\,e^{ik\left( {z - 
z'} \right)}\left[ {\sum\limits_{m = - \infty }^\infty {e^{im\left( {\varphi 
- \varphi '} \right)}I_m \left( {k\rho _ < } \right)K_m \left( {k\rho _ > } 
\right)} } \right]}, 
\end{equation}
\end{widetext}

\noindent
where $I_m ,K_m $ are the modified Bessel functions whose wronskian is  
$W\left\{ {I_m \left( \rho \right),K_m \left( \rho \right)} \right\} = - 1 
\mathord{\left/ {\vphantom {1 \rho }} \right. \kern-\nulldelimiterspace} 
\rho $ and $\rho _ < (\rho _ > )$ is the smaller (larger) of \textit{$\rho $} and \textit{$\rho $'}. It is easy 
to check that the eigenfunctions of $\hat {M}$ and of its adjoint are the 
same, namely $e^{ikz}e^{im\theta }$, while the eigenvalues are equal to

\begin{equation}
\label{eq35}
\chi _{km} = \frac{1}{2} - k\rho _0 I_m \left( {k\rho _0 } \right)'K_m 
\left( {k\rho _0 } \right).
\end{equation}

\noindent
If we consider $k = 0$ then

\begin{equation}
\label{eq36}
\chi _{00} = \frac{1}{2}
\end{equation}

\noindent
and

\begin{equation}
\label{eq37}
\chi _{0m} = 0
\end{equation}

\noindent
for $m \ne 0$ which are the eigenvalues of a disk in 2D \cite{Khavison2007}. $\hat {M}$ and 
its adjoint having the same eigenfunctions is a reminiscense of the fact 
that the electrostatic operator is symmetric for disks in 2D \cite{Ahner1986b}.

Both types of cylinders have a continuum spectrum, which is a signature of accommodating running wave 
along the z-direction, since cylindrical 
surfaces are unbounded hence, non-compact. Similar to phonon polaritons and 
implicitly using the separation of variables method the full spectrum of 
surface plasmons was calculated for spherical \cite{Prodan2004}, for spheroidal \cite{Brandl2007}, and 
for circular cylindrical shapes \cite{Morandi2008,Wan2013} in the so called hybridization model 
of plasmon resonances. Analysis of plasmon resonances in \cite{Prodan2004,Brandl2007,Morandi2008,Wan2013} leads to the same 
conclusion regarding the eigenvalues of the electrostatic operator.

All these shapes presented above generate also shelled configurations made 
of confocal surfaces determined by the above coordinate systems. These 
shelled systems support cavity and particle polaritons that interact among 
themselves. The interaction between cavity and particle modes however takes 
place in the subspace determined by each eigenvalue, thus it is possible to 
calculate all eigenmodes in shelled configurations. Calculations of shelled 
structures were performed in Refs. \cite{Englman1966,Englman1968,Prodan2004,Brandl2007,Morandi2008,Wan2013}. 

\section{Discussions}

\begin{figure}[!h]
\centering
\includegraphics [width=6in,height=4in] {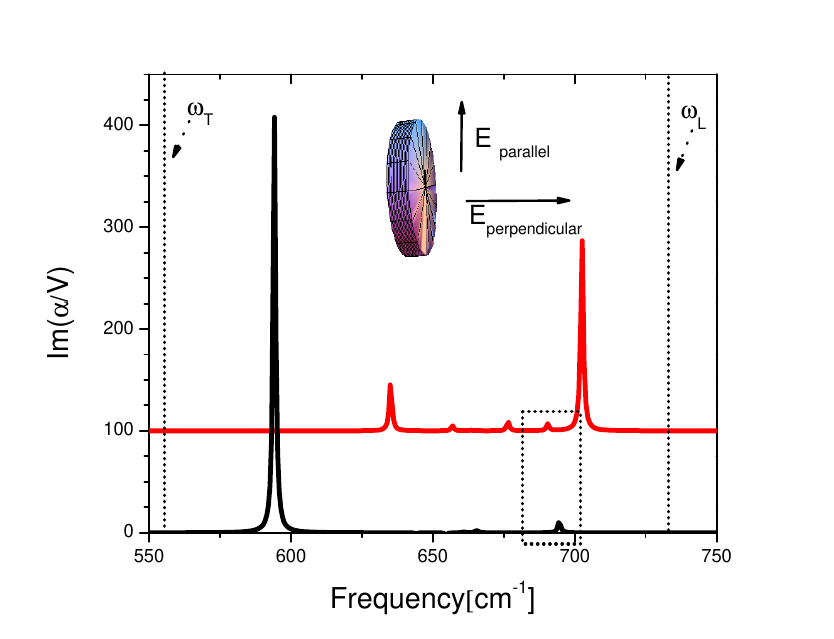}
\caption{\label{fig:2}
(Color online) The resonance modes of a GaN disk with $1/4.7$ aspect ratio for parallel and perpendicular polarizations. 
The modes for 
perpendicular polarization are shifted upwards for better clarity. The inset shows the disk 
and the two polarizations of the field. The dotted rectangle highlights those two resonances 
discussed in the main text and the two dotted vertical lines indicate $\omega_T$ and $\omega_L$, respectively.}
\end{figure}

The BIE method can be a method of choice in many situations even when 
nanoparticles are fabricated from crystals with an anisotropic dielectric 
constant. In this case the BIE method can be perfectly adapted by adopting 
the approach used in Ref. \cite{Fonoberov2004}. The BIE method can be also a convenient 
tool to comprehend and calibrate results with mixed states made by coupling 
continuum and localized states like those studied in a recent work \cite{Gubbin2016}. 
Moreover, a boundary integral equation method provides more insight about 
the localized resonance modes. For instance in Ref. \cite{Feng2015} there are studied the 
LSPhRs in an array of micro-disks of gallium nitride on silicon carbide 
substrate. The effect of localized surface phonon polariton resonances were 
studied by reflectance measurements in two polarization configurations, 
transverse electric (TE) and transverse magnetic (TM). The authors observed 
a little discrepancy between the measurements, which show only one resonance, and 
the finite element simulations, which predict two close resonances. We calculated the 
eigenvalues and their weights for a smooth but also a similar micro-disk that in turn has an 
aspect ratio of $1/4.7$. In the reflectance measurements in the TE configuration 
the electric field is parallel to the surface of the disk, while in the TM 
configuration both types of modes (i. e., parallel and perpendicular to the 
disk surface) are excited. Although our calculations are qualitative since 
they are made for a single disk some insight can be obtained since we have 
obtained all possible resonances of such micro-disk. In Fig.~\ref{fig:2} we may see 
that for the electric parallel to the disk surface there are only two 
resonances of which one is predominant. On the contrary, for the electric 
field perpendicular to the disk there are several resonances of which two 
are predominant. With the BIE method we may see that the small discrepancy 
noticed in Ref. \cite{Feng2015} is from two small resonances that come from different 
field polarizations. These two resonances are highlighted by a dotted 
rectangle.

Spectral studies of the electrostatic operator has sparked new interest 
arising from mathematical theory regarding plasmons, cloaking as well as 
various inverse problems \cite{Nicorovici1994,Ando2016,Ammari2014,Ammari2016,Ammari2013,Milton2005,Milton2006}. For bounded and smooth surfaces the 
electrostatic operator is compact with a countable number of eigenvalues 
that accumulates at the origin which belongs to the essential spectrum \cite{Khavison2007}. 
The eigenvalues lie within (-1/2,1/2] with the eigenvalue 1/2 corresponding 
to the equilibrium charge distribution \cite{Khavison2007,Sandu2013b}. For shapes resulted from 
separable coordinate systems, the capacitance of equilibrium charge can be calculated by simple integration \cite{Sandu2013b}.

Questions regarding spectral properties like the location of the essential 
spectrum of the electrostatic operator and the asymptotic form of its 
eigenfunctions turn out to be of great interest. Some form of cloaking is 
related to anomalous localized resonance that takes place at the 
accumulation point of the eigenvalues  \cite{Nicorovici1994,Ando2016,Ammari2013,Milton2005,Milton2006}. Several 2D systems 
have been considered for cloaking including concentric disks \cite{Nicorovici1994,Milton2005,Milton2006}, 
ellipses \cite{Ando2016}, and confocal ellipses \cite{Chung2014}. All these approaches need also 
the eigenfunctions beside all the eigenvalues. The results of previous 
section regarding the eigenvalues and the eigenfunctions of the electrostatic operator defined on different surfaces
and on their shelled structures counterparts may be used to extend the analysis of cloaking from disks and 
ellipses to 3D setting. 

\begin{figure}[h]
\centering
\includegraphics [width=6in,height=4in] {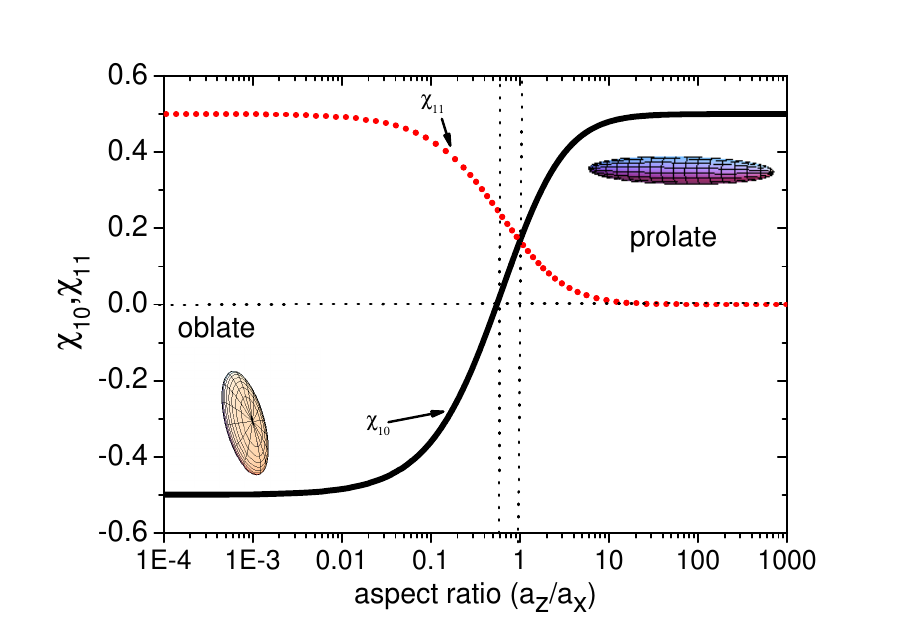}
\caption{\label{fig:3}
(Color online) The eigenvalues $\chi _{10}$ (black solid line) and $\chi _{11}$ (red dotted line) as a function of aspect ratio for prolate and oblate spheroids. One of the vertical dotted lines indicates the aspect ratio (an oblate) for which $\chi _{10} = 0$ and the other shows the aspect ratio for sphere where the $\chi _{10} = \chi _{11} = 1/6$.
}
\end{figure}

Since 0 belongs to the essential spectrum a legitimate question is if there 
are shapes which have 0 as an eigenvalue and more general if there is a 
shape that may have as an eigenvalue any number within -1/2 and 1/2. The general answer is given 
in \cite{Khavison2007} and in \cite{Ahner1994b} for spheroids and more recently in \cite{Feng2016} for 
general ellipsoids. In Fig.~\ref{fig:3} we present a graphical proof where we plotted 
the eigenvalues $\chi _{lm} $ with $l = 1$, $m = 0,1$ for prolate and oblate 
spheroids. We denote the ratio $a_{z}$/$a_{x}$  as the aspect ratio, where 
$a_{z}$ is the semi-axis along $z$-direction and $a_{x}$ is the semi-axis along 
$x$-direction. An aspect ratio $>1$ defines a prolate spheroid and an aspect 
ratio $<1$ defines an oblate spheroid. Prolate spheroids reach the value 1/2 
very fast at a rate inverse proportional to the square of the aspect ratio 
(i. e., $\left( {1 \mathord{\left/ {\vphantom {1 2}} \right. 
\kern-\nulldelimiterspace} 2 - \chi _{10} } \right) \sim \left( {{a_x } 
\mathord{\left/ {\vphantom {{a_x } {a_z }}} \right. 
\kern-\nulldelimiterspace} {a_z }} \right)^2)$ thus, practically, for aspect 
ratios $>10$, $\chi _{10} $ is pretty close to 1/2. From Eqs. (\ref{eq7}) and (\ref{eq10}) we can see that 
this eigenvalue generates a resonance that is redshifted to IR (theoretically it goes to 0) for plasmons 
or is moved toward $\omega _T$ for phonon polaritons. 

It can be seen in Fig.~\ref{fig:3} that there is an oblate spheroid with an aspect 
ratio between 0.1 and 1 for which $\chi _{10} = 0$. Furthermore, as the 
aspect ratio tends to 0 $\chi _{10} $ approaches -1/2 as $\left( {1 
\mathord{\left/ {\vphantom {1 2}} \right. \kern-\nulldelimiterspace} 2 + 
\chi _{10} } \right) \sim \left( {{a_z } \mathord{\left/ {\vphantom {{a_z } 
{a_x }}} \right. \kern-\nulldelimiterspace} {a_x }} \right)^2$ and $\chi 
_{11} $ goes to 1/2 as $\left( {1 \mathord{\left/ {\vphantom {1 2}} \right. 
\kern-\nulldelimiterspace} 2 - \chi _{11} } \right) \sim \left( {{a_z } 
\mathord{\left/ {\vphantom {{a_z } {a_x }}} \right. 
\kern-\nulldelimiterspace} {a_x }} \right)^2$. Similarly, from Eqs. (\ref{eq7}) and (\ref{eq10}) 
we see that if the electrostatic operator has an eigenvalues close to -1/2 the 
plasmon resonance frequency approaches the plasma frequency of the bulk 
material and the phonon polariton resonances shifts toward $\omega _L $. 

In other words, asymptotically for $l$=1 a 2D disk embedded in 3D has an 
eigenvalue -1/2 for fields perpendicular to the surface of the disk and an 
eigenvalue equals to 1/2 for fields parallel to the surface. The rates of 
approaching these values are proportional to the square of the aspect ratio. 
On the other hand, in contrast to oblate disks, for finite and constant-thickness 
disks, like those discussed at the beginning of this section, 
there is an eigenvalue that approaches 1/2 at a rate proportional to the aspect ratio, 
which is defined as the ratio of the thickness and disk diameter. This 
assertion is shown in the following. This mode defines the 
graphene plasmon resonance which is excited by an electric field parallel to 
the graphene plane \cite{Fang2013}. For instance, considering graphene as a 2D sheet, 
the graphene plasmon frequency of nano-disks scales as $1 \mathord{\left/ 
{\vphantom {1 {\sqrt d }}} \right. \kern-\nulldelimiterspace} {\sqrt d }$, 
where $d$ is the diameter of of graphene nano-disk \cite{Fang2013}. In 3D models, 
graphene nano-disks 
are simulated as finite-thickness disks with a bulk complex conductivity 
obtained from graphene sheet-conductivity divided by the disk thickness $h$. 
Hence the finite thickness nano-disk has the following dielectric relative 
permittivity \cite{Vakil2011}

\begin{equation}
\label{eq38}
\varepsilon _d \left( \omega \right) = 1 + \frac{i\sigma \left( \omega \right)} 
{\varepsilon _v \omega h}.
\end{equation}
Here $\sigma \left( \omega \right) = {ie^2E_F } \mathord{\left/ {\vphantom 
{{ie^2E_F } {\left[ {\left( {\pi \hbar ^2} \right)\left( {\omega + i 
\mathord{\left/ {\vphantom {i \tau }} \right. \kern-\nulldelimiterspace} 
\tau } \right)} \right]}}} \right. \kern-\nulldelimiterspace} {\left[ 
{\left( {\pi \hbar ^2} \right)\left( {\omega + i \mathord{\left/ {\vphantom 
{i \tau }} \right. \kern-\nulldelimiterspace} \tau } \right)} \right]}$ is 
the graphene sheet conductivity that has a Drude-like form and $\varepsilon 
_v $ is the vacuum permittivity. We denoted by $E_{F}$ the Fermi energy in 
graphene disk, \textit{$\tau $} is the relaxation time, and $\hbar$ is the reduced Planck constant. 
Thus, the plasma frequency for a 
3D model of graphene takes the form $\omega _p^2 = {e^2E_F } \mathord{\left/ 
{\vphantom {{e^2E_F } {\left( {\pi \hbar ^2\varepsilon _v } \right)}}} 
\right. \kern-\nulldelimiterspace} {\left( {\pi \hbar ^2\varepsilon _v h} 
\right)}$. Bearing in mind that the eigenvalues are scale invariant they are also a function of 
the aspect ratio $h$/$d$ only. Since the plasmon frequency is $\tilde {\omega 
}_{pk} = \omega _p \sqrt {1 / 2 - \chi _k } $ (see Eq. (\ref{eq7}))  we may 
deduce that the graphene 
plasmon frequency scales as $1 \mathord{\left/ {\vphantom {1 {\sqrt d }}} 
\right. \kern-\nulldelimiterspace} {\sqrt d }$ and that 
its corrsponding eigenvalue goes to $1/2$ at the rate linear with $h$/$d$ and not quadratically like 
in the case of oblate disks. 

Owing to the success of the BIE in the quasi-static regime there is a renewed 
interest in the modal approach based on BIE to plasmon problems in the 
full-wave electromagnetic regime \cite{Makitalo2014}. The full-wave electromagnetic 
treatment discussed in Ref. \cite{Makitalo2014} has at its core the free-space Green's function of 
scalar Helmholtz equation that is also separable in those eleven coordinate system in which 
Laplace eqation separates \cite{Boyer1976,Miller1977}. In general, in the full-wave regime, due to 
radiation losses, the eigenvalues are no longer real as in the case of 
quasistatic regime \cite{Makitalo2014}. It will be interesting to apply the same symmetry 
arguments about Helmholtz equation to the spectrum of the boundary integral 
operators used in the full-wave treatment for the shapes and their 
corresponding shells discussed in this work. In this way one can study 
optical bound states \cite{Hsu2016} with no radiation losses in structures having other forms than 
spherical shapes which have been considered recently \cite{Silveirinha2014,Monticone2014}.

\section{Conclusions}

In the present work we systematically apply the boundary integral equation 
method to obtain explicit relations between the eigenvalues of the 
electrostatic operator associated with the method and the localized surface 
phonon/plasmon polariton resonances. The boundary integral equation method 
shows real benefits by explicitly assigning modes, spectral features, and resonances 
to experimental data. Active modes of thin disks are analysed asymptotically with respect to disk 
thickness. Two examples were considered: the localized surface phonon 
resonances in GaN micro-disks and the scaling of localized surface plasmon 
resonance frequency with the size of graphene nano-disks. 

We have shown that, regarding the spectra of both surface phonons and 
plasmons, all the calculations performed so far in separable coordinates 
contain not only all the eigenvalues but also all the eigenfunctions of the 
electrostatic operators. 

Group-theoretical analysis applied to Laplace 
equation may be exploited in obtaining the eigenvalues and eigenvectors of 
the electrostatic operator for shapes and their corresponding shell 
structures described by separable coordinate systems. We calculated explicitly the 
eigenvalues and the eigenfunctions of elliptic and circular cylinders, which with the known 
results about other shapes may be 
used to extend the study of cloaking by 
anomalous localized resonance from 2D to 3D structures. 
The results of Lie group analysis may be further employed to study the spectrum of boundary operators 
associated with the full-wave regime.The extension to the full-wave regime creates the opportunity of studying and obtaining bound states in continuum with shapes different from spherical geometry.

\begin{acknowledgments}
The work was supported by a grant of the Romanian National Authority for 
Scientific Research, CNCS-UEFISCDI, Project Number 
PNII-ID-PCCE-2011-2-0069.

\end{acknowledgments}


\end{document}